# Limited Memory Prediction for Linear Systems with Different types of Observation

Ha-ryong Song, Vladimir Shin

**Abstract**— This paper is concerned with distributed limited memory prediction for continuous-time linear stochastic systems with multiple sensors. A distributed fusion with the weighted sum structure is applied to the optimal local limited memory predictors. The distributed prediction algorithm represents the optimal linear fusion by weighting matrices under the minimum mean square criterion. The algorithm has the parallel structure and allows parallel processing of observations making it reliable since the rest faultless sensors can continue to the fusion estimation if some sensors occur faulty. The derivation of equations for error cross-covariances between the local predictors is the key of this paper. Example demonstrates effectiveness of the distributed limited memory predictor.

——————————— ◆ ———————————

## 1 INTRODUCTION

IN many applications, multiple sensors observe a common origin, for example, system state [1, 2]. If there are no constraints on communication channels and processor bandwidths, complete measurements can be brought to a central processor for data processing. In this case, sensors act as simple data collectors and do not perform any data processing. One of the advantages of this centralized processing is that it involves minimal information loss. However, it can result in severe computational overhead. Also, there needs to be a somewhat ideal assumption on the environment mentioned above.

In practice, especially when sensors are dispersed over a wide geographic area, there are limitations on the amount of communications allowed among sensors. Also, sensors are provided with processing capabilities. In this case, a certain amount of computation can be performed at the individual sensors and a compressed version of sensor data can be transmitted to a fusion center where the received information is appropriately combined to yield the global inference. The advantage of the distribution of filters is that the parallel structures would lead to increase the input data rates and make easy fault detection and isolation. However, the accuracy of the distributed filter or predictor is generally lower than of the centralized filter. Various distributed and parallel versions of the standard Kalman filter have been reported for linear dynamic systems [2-8].

To get a more accurate estimate of a state of system under potential uncertainty, various kinds of techniques have been introduced and discussed. Among them, a limited memory technique called receding horizon, which is robust against temporal uncertainty, has been rigorously investigated [9-12]. In this paper, we study the distributed receding horizon fusion prediction for the continuous-time linear systems with multiple sensors. The local re-ceding horizon predictors (LRHPs), which we fuse, utilize finite measurements over the most recent time interval [9-12]. It has been a general rule that the LRHPs are often robust against dynamic model uncertainties and numerical errors than the standard filters, which utilize all measurements. Based on the LRHPs [10, 12] and the optimal fusion formula with matrix weights [8, 13], we propose a distributed receding horizon predictor which has a better accuracy than every LRHP, and it has the reduced computational burden as compared to the centralized receding horizon predictor (CRHP).

This paper is organized as follows. The problem is set up in Section II. The CRHP is described in Section III. In Section IV, we present the main result regarding the distributed receding horizon predictor (DRHP) for multisensor environment. Here the key equations for cross-covariances between LRHPs are derived. In Section V, example illustrates performance of the CRHP and DRHP. In Section VI, concluding remarks are given

## 2 STATEMENT OF PROBLEM

Consider the continuous-time linear stochastic system with sensors

$$\dot{x}_t = F_t x_t + G_t v_t, \quad t \geq t_0, \quad x_0 = x_{t_0}, \tag{1}$$

$$y_t^{(i)} = H_t^{(i)} x_t + w_t^{(i)}, \quad i = 1,\ldots,N, \tag{2}$$

where $x_t \in \Re^n$ is the state, $y_t^{(i)} \in \Re^{m_i}$ is the measurement, $v_t \in \Re^r$ and $w_t^{(i)} \in \Re^{m_i}$ are uncorrelated white Gaussian noises with zero mean and intensity matrices $Q_t$ and $R_t^{(i)}$, respectively, and $F_t$, $G_t$, $H_t^{(i)}$ are matrices with compatible dimensions. Superscript $i$ denotes the $i$ th sensor, $N$ is the number of sensors.

The initial state $x_0 \sim N(\bar{x}_0, P_0)$, $\bar{x}_0 = E(x_0)$, $P_0 = \text{cov}(x_0, x_0)$ is assumed to be uncorrelated of $v_t \in \Re^r$ and $w_t^{(i)} \in \Re^{m_i}$, $i = 1,\ldots,N$.

Our purpose is to find the distributed weighted fusion predictor $\hat{x}_{t+\Delta}, \Delta \geq 0$ of the state $x_{t+\Delta}$ based on the over-

————————————————
• Ha-ryong Song and Vladimir Shin are with the Gwangju Institute of Science and Technology, 216 Cheomdan-gwagiro (Oryong-dong), Buk-Gu, Gwangju, South Korea.

all horizon sensor measurements

$$Y_{t-T}^t = \{y_s^{(1)},...,y_s^{(N)}, \ t-T \leq s \leq t\}. \tag{3}$$

## 3 CENTRALIZED LIMITED MEMORY PREDICTOR

Let consider the original dynamic system (1) and rewrite the measurement model (2) in equivalent form. We obtain

$$\dot{x}_t = F_t x_t + G_t v_t, \ t \geq t_0, \ x_0 = x_{t_0},$$
$$Y_t = H_t x_t + w_t, \tag{4}$$

where

$$Y_t = \begin{bmatrix} y_t^{(1)} \\ \vdots \\ y_t^{(N)} \end{bmatrix}, \ H_t = \begin{bmatrix} H_t^{(1)} \\ \vdots \\ H_t^{(N)} \end{bmatrix}, \ w_t = \begin{bmatrix} w_t^{(1)} \\ \vdots \\ w_t^{(N)} \end{bmatrix}. \tag{5}$$

Then the optimal *centralized fusion receding horizon predictor* (CRHP) $\hat{x}_{t+\Delta}^{opt} \equiv \hat{x}_{t+\Delta}^{CRHP}$ and its error covariance $P_{t+\Delta}^{opt} \equiv P_{t+\Delta}^{CRHP}$ are determined by the Kalman predictor equations [14]:

$$\dot{\hat{x}}_\tau^{CRHP} = F_\tau \hat{x}_\tau^{CRHP},$$
$$\dot{P}_\tau^{CRHP} = F_\tau P_\tau^{CRHP} + P_\tau^{CRHP} F_\tau' + \tilde{Q}_\tau, \tag{6}$$
$$\tilde{Q}_\tau = G_\tau Q_\tau G_\tau', t \leq \tau \leq t+\Delta,$$

where the superscript / denotes transpose, and the current initial conditions at time $\tau = t$,

$$\hat{x}_{\tau=t}^{CRHP} = \hat{x}_t^{CRHF}, P_{\tau=t}^{CRHP} = P_t^{CRHF}, \tag{7}$$

are given by the *centralized receding horizon filter* (CRHF) equations [10-12]:

$$\dot{\hat{x}}_s^{CRHF} = F_s \hat{x}_s^{CRHF} + L_s^{CRHF}\left[Y_s - H_s \hat{x}_s^{CRHF}\right],$$
$$L_s^{CRHF} = P_s^{CRHF} H_s' R_s^{-1},$$
$$\dot{P}_s^{CRHF} = F_s P_s^{CRHF} + P_s^{CRHF} F_s' + \tilde{Q}_s$$
$$\quad - P_s^{CRHF} H_s' R_s^{-1} H_s P_s^{CRHF}, \tag{8}$$
$$R_s = diag\left[R_s^{(1)},...,R_s^{(N)}\right],$$
$$P_s^{CRHF} = E\left[(x_s - \hat{x}_s^{CRHF})(x_s - \hat{x}_s^{CRHF})'\right],$$
$$t-T \leq s \leq t.$$

The horizon initial conditions at time instant $s = t-T$ in (8) represent the unconditional mean and covariance of the horizon state $x_{t-T}$, i.e.,

$$\hat{x}_{s=t-T}^{CRHF} = E(x_{t-T}) \equiv \bar{x}_{t-T},$$
$$P_{s=t-T}^{CRHF} = E\left[(x_{t-T} - \bar{x}_{t-T})(x_{t-T} - \bar{x}_{t-T})'\right] = P_{t-T}, \tag{9}$$

which satisfy the Lyapunov differential equations

$$\dot{\bar{x}}_\tau = F_\tau \bar{x}_\tau, \ \bar{x}_{t_0} = \bar{x}_0,$$
$$\dot{P}_\tau = F_\tau P_\tau + P_\tau F_\tau' + \tilde{Q}_\tau, \ P_{t_0} = P_0, \tag{10}$$
$$t_0 \leq \tau \leq t-T,$$

where the initial mean $\bar{x}_0$ and covariance $P_0$ are given.

Let us now summarize the algorithmic procedure for the CRHP. First, the horizon initial conditions (9) $\hat{x}_{t-T}^{CRHF} \equiv \bar{x}_{t-T}$ and $P_{t-T}^{CRHF} \equiv P_{t-T}$ are calculated by solution of the Lyapunov equations (10). Second, the CRHF equations (8) for the current initial conditions (7) are solved on the horizon interval $s \in [t-T;t]$ using the current horizon measurements $Y_s$ and finally, the CRHP $\hat{x}_{t+\Delta}^{CRHP}$ is calculated by (6).

The CRHP represents a joint estimator. To predict the state estimate $\hat{x}_{t+\Delta}^{CRHP}$, the implementation of the CRHF (8) requires all the horizon sensor measurements (5) jointly at each time instant $t$. Therefore, in the case of several limitations, such as computational cost, communication resources, the CRHF and CRHP cannot produce well-timed results, especially for the large number of sensors. So the distributed receding horizon predictor is preferable as there is no need to predict state vector by using *overall* sensor measurements (5) simultaneously.

## 3 DISTRIBUTED LIMITED MEMORY PREDICTOR

Now we show that the fusion formula [8, 13] can serve as the basis for designing of a distributed predictor. A new *distributed receding horizon predictor* (DRHP) is described as follows: first, the local sensor measurements $y_s^{(1)},...,y_s^{(N)}$ are processed separately by applying the receding horizon predictor (6)-(10) to the following dynamic subsystem:

$$\dot{x}_t = F_t x_t + G_t v_t, \ t \geq t_0, \ x_0 = x_{t_0},$$
$$y_t^{(i)} = H_t^{(i)} x_t + w_t^{(i)}, \ (index \ i \ is \ fixed), \tag{11}$$

and second, the obtained local receding horizon predictors are fused in an optimal linear combination.

### 3.1 Local Limited memory Filters and Predictors

Let denote the local receding horizon predictor of the state $x_{t+\Delta}$ based on the individual receding horizon sensor measurements $\{y_s^{(i)}, t-T \leq s \leq t\}$ by $\hat{x}_{t+\Delta}^{(i)}$. Then the local receding horizon prediction $\hat{x}_{t+\Delta}^{(i)}$ and filtering $\hat{x}_t^{(i)}$ estimates, and corresponding error covariances $P_{t+\Delta}^{(ii)}$ and $P_t^{(ii)}$ are determined by the analogous equations (6) and (8), (10), respectively. We have

$$\dot{\hat{x}}_\tau^{(i)} = F_\tau \hat{x}_\tau^{(i)},$$
$$\dot{P}_\tau^{(ii)} = F_\tau P_\tau^{(ii)} + P_\tau^{(ii)} F_\tau' + \tilde{Q}_\tau, \quad (6')$$
$$t \leq \tau \leq t + \Delta.$$

$$\dot{\hat{x}}_s^{(i)} = F_s \hat{x}_s^{(i)} + L_s^{(i)} \left[ y_s^{(i)} - H_s \hat{x}_s^{(i)} \right], \quad t-T \leq s \leq t,$$
$$L_s^{(i)} = P_s^{(ii)} H_s^{(i)'} R_s^{(i)-1},$$
$$\dot{P}_s^{(ii)} = F_s P_s^{(ii)} + P_s^{(ii)} F_s' + \tilde{Q}_s$$
$$\quad - P_s^{(ii)} H_s^{(i)'} R_s^{(i)-1} H_s^{(i)} P_s^{(ii)}, \quad (8')$$
$$P_s^{(ii)} = E\left[ (x_s - \hat{x}_s^{(i)})(x_s - \hat{x}_s^{(i)})' \right],$$
$$\hat{x}_{t-T}^{(i)} = \overline{x}_{t-T}, \quad P_{t-T}^{(ii)} = P_{t-T}.$$

$$\dot{\overline{x}}_\tau = F_\tau \overline{x}_\tau, \quad \overline{x}_{t_0} = \overline{x}_0,$$
$$\dot{P}_\tau = F_\tau P_\tau + P_\tau F_\tau' + \tilde{Q}_\tau, \quad P_{t_0} = P_0, \quad (10')$$
$$t_0 \leq \tau \leq t - T.$$

Thus, after sequentially solution of the equations (10′), (8′), and (6′) we have $N$ local receding horizon prediction estimates,

$$\hat{x}_{t+\Delta}^{(1)}, \ldots, \hat{x}_{t+\Delta}^{(N)}, \quad (12)$$

and corresponding error covariances,

$$P_{t+\Delta}^{(11)}, \ldots, P_{t+\Delta}^{(NN)}. \quad (13)$$

Using these values we propose new distributed prediction algorithm.

### 3.2 Distributed Limited memory Predictor

The distributed receding horizon estimate $\hat{x}_{t+\Delta}^{DRHP}$ based on the *overall* sensor measurements (3) is constructed by using the fusion formula, i.e.,

$$\hat{x}_{t+\Delta}^{DRHP} = \sum_{i=1}^{N} W_{t,\Delta}^{(i)} \hat{x}_{t+\Delta}^{(i)}, \quad \sum_{i=1}^{N} W_{t,\Delta}^{(i)} = I_n, \quad (14)$$

where $I_n$ is the identity matrix, and $W_{t,\Delta}^{(1)}, \ldots, W_{t,\Delta}^{(N)}$ are the time-varying weighted matrices determined by the mean square criterion.

**Theorem 1** (a) *The optimal weights* $W_{t,\Delta}^{(1)}, \ldots, W_{t,\Delta}^{(N)}$ *satisfy the linear algebraic equations* [8, 13]:

$$\sum_{i=1}^{N} W_{t,\Delta}^{(i)} \left[ P_{t+\Delta}^{(ij)} - P_{t+\Delta}^{(iN)} \right] = 0, \quad \sum_{i=1}^{N} W_{t,\Delta}^{(i)} = I_n,$$
$$j = 1, \ldots, N-1, \quad (15)$$

and they can be explicitly written in the following form

$$W_{t,\Delta}^{(i)} = \sum_{j=1}^{N} D_{t+\Delta}^{(ij)} \left( \sum_{l,h=1}^{N} D_{t+\Delta}^{(lh)} \right)^{-1}, \quad i = 1, \ldots N, \quad (16)$$

where $D_{t+\Delta}^{(ij)}$ is the (ij)th $n \times n$ submatrix of the $nN \times nN$ block matrix $P_{t+\Delta}^{-1}$, $P_{t+\Delta} = \left[ P_{t+\Delta}^{(ij)} \right]_{i,j=1}^{N}$.

(b) *The distributed error covariance*

$$P_{t+\Delta}^{DRHP} = E\left[ (x_{t+\Delta} - \hat{x}_{t+\Delta}^{DRHP})(x_{t+\Delta} - \hat{x}_{t+\Delta}^{DRHP})' \right], \quad (17)$$

*is given by*

$$P_{t+\Delta}^{DRHP} = \sum_{i,j=1}^{N} W_{t,\Delta}^{(i)} P_{t+\Delta}^{(ij)} W_{t,\Delta}^{(j)'}. \quad (18)$$

Equations (14)-(18) defining the unknown weights $W_{t,\Delta}^{(i)}$, $i=1,\ldots,N$ and distributed error covariance $P_{t+\Delta}^{DRHP}$ depend on the local covariances (13) $P_{t+\Delta}^{(ii)}, i=1,\ldots,N$, which are determined by (6′), (8′), (10′), and local cross-covariances

$$P_{t+\Delta}^{(ij)} = E\left[ (x_{t+\Delta} - \hat{x}_{t+\Delta}^{(i)})(x_{t+\Delta} - \hat{x}_{t+\Delta}^{(j)})' \right],$$
$$i, j = 1, \ldots, N, i \neq j. \quad (19)$$

Equations for local cross-covariances (19) we derive in Theorem 2.

**Theorem 2** *The local cross-covariances (19) satisfy the following differential equations:*

$$\dot{P}_\tau^{(ij)} = F_\tau P_\tau^{(ij)} + P_\tau^{(ij)} F_\tau', \quad t \leq \tau \leq t + \Delta, \quad (20)$$

*with the current initial conditions* $P_{\tau=t}^{(ij)} = P_\tau^{(ij)}$ *determining by*

$$\dot{P}_s^{(ij)} = \tilde{F}_s^{(i)} P_s^{(ij)} + P_s^{(ij)} \tilde{F}_s^{(j)'} + \tilde{Q}_s,$$
$$\tilde{F}_s^{(i)} = F_s - L_s^{(i)} H_s^{(i)}, i, j = 1, \ldots, N, i \neq j, \quad (21)$$
$$t-T \leq s \leq t,$$

*where the horizon initial conditions*

$$P_{s=t-T}^{(ij)} = E\left[ (x_{t-T} - \hat{x}_{t-T}^{(i)})(x_{t-T} - \hat{x}_{t-T}^{(j)})' \right]$$
$$= E\left[ (x_{t-T} - \overline{x}_{t-T}^{(i)})(x_{t-T} - \overline{x}_{t-T}^{(j)})' \right] = P_{t-T}, \quad (22)$$

*and filter gains* $L_s^{(i)}$ *determined by (10′) and (8′), respectively.*

The derivation of equations (20) and (21) is given in Appendix.

Thus, equations (14)-(22) completely define the DRHP.

*Remark 1.* If the prediction time $\Delta = 0$ the equations (14)-(22) determine the distributed receding horizon filter.

*Remark 2.* The local receding horizon estimates $\hat{x}_t^{(i)}$, $\hat{x}_{t+\Delta}^{(i)}$, $i = 1,\ldots,N$ are separated for different types of sensors, i.e., each local estimate $\hat{x}_t^{(i)}$ (or $\hat{x}_{t+\Delta}^{(i)}$) is found independently of other estimates. Therefore, the local estimates can be implemented in parallel for different sensors (2).

*Remark 3.* We may note that the local error cross-covariances $P_t^{(ij)}$, $P_{t+\Delta}^{(ij)}$ and the weights $W_{t,\Delta}^{(i)}$ may be pre-computed, since they do not depend on the sensor measurements, but only on the noise statistics $Q_t$ and $R_t^{(i)}$ and the system matrices $F_t$, $G_t$, $H_t^{(i)}$ which are the part of system model (1), (2). Thus, once the measurement schedule has been settled, the real-time implementation of the DRHP requires only the computation of the local prediction estimates $\hat{x}_{t+\Delta}^{(i)}$, $i = 1,\ldots,N$ and the final distributed estimate $\hat{x}_{t+\Delta}^{DRHP}$.

## 4 EXAMPLE

### 4.1 Multisensor system model

Here we verify the CRHP and DRHP using a linearized model of the water tank system taken from [15]. We have

$$\dot{x}_t = \begin{bmatrix} -0.0139 & 0 & 0 \\ 0 & -0.0277 & 0 \\ 0 & 0.1667 & -0.1667 + 0.1\delta_t \end{bmatrix} x_t + \begin{bmatrix} 1 \\ 1 \\ 1 \end{bmatrix} v_t, \quad t \geq 0, \quad (23)$$

where $x_t = [x_{1,t}, x_{2,t}, x_{3,t}]'$, $x_{1,t}$ is the water level, $x_{2,t}$ is water temperature and $x_{3,t}$ is sensor's own temperature, $\delta_t$ is an uncertain model parameter, $v_t$ is a white Gaussian noise. The system noise intensity $Q_t$ is $0.02^2$ and $\delta_t = 1$ on the interval $1 \leq t \leq 3$.

The measurement model contains four identical local sensors, one of which is main while others are reserved. We have

$$y_t^{(i)} = H^{(i)} x_t + w_t^{(i)}, \quad H^{(i)} = \begin{bmatrix} 0 & 0 & 1 \end{bmatrix}, \quad i = 1,\ldots,4, \quad (24)$$

where $w_t^{(i)}$, $i = 1,\ldots,4$ are white Gaussian noises with intensities $R_t^{(i)} = 0.01^2$, $i = 1,\ldots,4$.

The horizon length and prediction time are taken as $T = 0.8$ and $\Delta = 0.5$, respectively.

Two predictors the CRHP ($\hat{x}_{t+\Delta}^{CRHP}$, $P_{t+\Delta}^{CRHP}$) and DRHP ($\hat{x}_{t+\Delta}^{DRHP}$, $P_{t+\Delta}^{DRHP}$) for the system model (23), (24) are compared. Generally, the distributed estimator is globally suboptimal compared with optimal centralized estimator [16]. Nevertheless, though accuracy of the distributed estimator may not be optimal, it has advantages of lower computational requirements, efficient communication costs, parallel implementation, and fault-tolerance [16–18]. Especially, in this paper, we show fault-tolerance of the DRHP. For this purpose, we propose the special simulation scenario where the sensors are sequentially breaking down:

(i) 4 sensors are working at $t \in \sum_1 = [0; 1.5]$;

(ii) 3 sensors are working at $t \in \sum_2 = (1.5; 2.5]$;

(iii) 2 sensors are working at $t \in \sum_3 = (2.5; 3.5]$;

(iv) 1 sensor is working at $t \in \sum_4 = (3.5; 5]$.

Note that if at least one of the sensors fails then the CRHP does not work, because the centralized predictor utilizes all sensor measurements (5). If one of the reserved sensors misses measurement data or fails then equations (8) can not be calculated. In this case the DRHP is still successfully working.

### 4.2 Results and Analysis

We focus on the analysis of the mean square error (MSE)

$$P_{33,t+\Delta} = E\left[\left(x_{3,t+\Delta} - \hat{x}_{3,t+\Delta}\right)^2\right], \quad \hat{x}_{3,t+\Delta} = \hat{x}_{3,t+\Delta}^{CRHP} \text{ or } \hat{x}_{3,t+\Delta}^{DRHP} \quad (25)$$

of the sensor temperature $x_{3,t}$ which directly contains the uncertainty $\delta_t$ in (23).

Simulation results for other coordinates are similar. The actual MSEs are calculated using the Monte-Carlo method with 1000 runs. Fig. 1 illustrates the time histories of the prediction estimates $\hat{x}_{3,t+\Delta}^{CRHP}$ and $\hat{x}_{3,t+\Delta}^{DRHP}$, and corresponding MSEs $P_{33,t+\Delta}^{CRHP}$ and $P_{33,t+\Delta}^{DRHP}$, respectively. As we see on Fig. 1, within the first time interval $\Sigma_1$ all sensors are working well and the CRHP is little better than the DRHP. However, later on when the sensors are sequentially breaking down the CRHP does not work on the intervals $\Sigma_2$, $\Sigma_3$ and $\Sigma_4$. Within these intervals the DRHP continues to work but its prediction accuracy ($P_{33,t+\Delta}^{DRHP}$) decreases as the sensors are breaking down. Also we observe that within the uncertainty interval $t \in [0; 1.5] \subset \Sigma_1$ the DRHP demonstrates good robust properties. The difference between MSEs $P_{33,t+\Delta}^{CRHP}$ and $P_{33,t+\Delta}^{DRHP}$ is negligible.

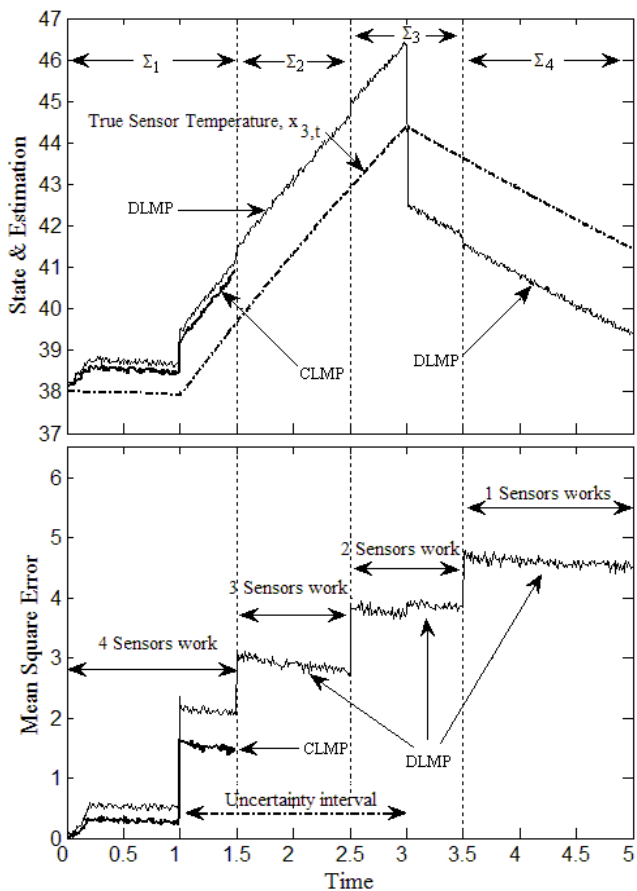

Fig 1. Comparison of CRHP and DRHP on different time intervals $\Sigma_i$, $i = 1,...,4$

Summarizing the simulation results we can conclude the DRHP can produce good and robust results.

## 5 CONCLUSIONS

In this paper we propose new DRHP for linear continuous-time dynamic systems with multisensor environment. It represents the weighted sum of the LRHPs. Each local predictor is fused by the minimum mean square error criterion. The matrix weights depend on cross-covariances between the local prediction errors. The differential equations for them are derived.

Furthermore, the DRHP has the parallel structure and allows parallel processing of observations making it reliable since the rest faultless sensors can continue to the fusion estimation if some sensors occur faulty. Simulation analysis and comparison with the CRHP verifies the effectiveness of the proposed distributed predictor.

## APPENDIX : DERIVATION OF EQUATIONS (20), (21)

According to (6') and (8') the local estimation errors $\tilde{x}_s^{(i)}$ and $\tilde{x}_s^{(j)}$, $i \neq j$ satisfy the following linear stochastic differential equations:

$$\dot{\tilde{x}}_s^{(i)} = \begin{cases} F_s^{(i)} \tilde{x}_s^{(i)}, & \text{for } t < s \leq t + \Delta, \\ \tilde{F}_s^{(i)} \tilde{x}_s^{(i)} + B_s^{(i)} \xi_s, & \text{for } t - T < s \leq t, \end{cases} \quad (A.1)$$

$$\dot{\tilde{x}}_s^{(j)} = \begin{cases} F_s^{(j)} \tilde{x}_s^{(j)}, & \text{for } t < s \leq t + \Delta, \\ \tilde{F}_s^{(j)} \tilde{x}_s^{(j)} + B_s^{(j)} \xi_s, & \text{for } t - T < s \leq t, \end{cases} \quad (A.2)$$

where

$$B_s^{(i)} = \begin{bmatrix} -L_s^{(i)} & 0 & G_s \end{bmatrix}, \; B_s^{(j)} = \begin{bmatrix} 0 & -L_s^{(j)} & G_s \end{bmatrix}, \quad (A.3)$$

and $\xi_s = \begin{bmatrix} w_s^{(i)'} & w_s^{(j)'} & v_s' \end{bmatrix}' \in \Re^{m_i + m_j + r}$ is the composite white Gaussian noise with the block intensity matrix

$$Q_s^{(\xi)} = diag \begin{bmatrix} R_s^{(i)} & R_s^{(j)} & Q_s \end{bmatrix}. \quad (A.4)$$

Then the cross-covariance $P_s^{(ij)}$ represents expectation of the product $E(\tilde{x}_s^{(i)} \tilde{x}_s^{(j)'})$ satisfying the Lyapunov differential equation [19]:

$$\dot{P}_s^{(ij)} = \frac{d}{ds} E \left[ \tilde{x}_s^{(i)} \tilde{x}_s^{(j)'} \right] = E \left[ \dot{\tilde{x}}_s^{(i)} \tilde{x}_s^{(j)'} + \tilde{x}_s^{(i)} \dot{\tilde{x}}_s^{(j)'} \right]$$
$$= \begin{cases} F_s P_s^{(ij)} + P_s^{(ij)} F_s', & t < s \leq t + \Delta, \\ \tilde{F}_s^{(i)} P_s^{(ij)} + P_s^{(ij)} \tilde{F}_s^{(i)'} + B_s^{(i)} Q_s^{\xi} B_s^{(j)'}, & t - T \leq s \leq t. \end{cases} \quad (A.5)$$

Since $B_s^{(i)} Q_s^{\xi} B_s^{(j)'} = \tilde{Q}_s$, we obtain (20) and (21).

This completes the proof of Theorem 2.

**Ha-ryong Song** He received the B.S. degree in Control and Instru-mentation Engineering from the Chosun University, Korea, in 2006, the M.S. degree in School of Information and Mechtronics from the Gwangju Institute of Science and Technology, Korea, in 2007. He is currently a Ph.D candidate in Gwangju Institute of Science and Technology. His research interests include estima-tion, target track-ing systems, data fusion, nonlinear filtering.

**Vladimir Shin** He received the B.Sc. and M.Sc. degrees in applied mathematics from Moscow State Aviation Institute, in 1977 and 1979, respectively. In 1985 he received the Ph.D. degree in mathe-matics at the Institute of Control Science, Russian Acad-emy of Sciences, Moscow. He is currently an Associate Profes-sor at Gwangju Institute of Science and Technology, South Korea. His research interests include estimation, filtering, tracking, data fusion, stochastic control, identification, and other multidi-mensional data processing methods.